\documentclass[aps,preprint,showpacs,showkeys]{revtex4}

\begin{document}

%%%%%%%%%%%%%%%%%%%%%% DEFINITIONS %%%%%%%%%%%%%%%%%%%%%%%%%%%%

\newcommand\beq{\begin{equation}}
\newcommand\eeq{\end{equation}}
\newcommand\beqa{\begin{eqnarray}}
\newcommand\eeqa{\end{eqnarray}}
\newcommand\ket[1]{|#1\rangle}
\newcommand\bra[1]{\langle#1|}
\newcommand\scalar[2]{\langle#1|#2\rangle}

\newcommand\jo[3]{\textbf{#1}, #3 (#2)}

%%%%%%%%%%%%%%%%%%%%%% TITLE PAGE %%%%%%%%%%%%%%%%%%%%%%%%%%%%%

\title{\Large\textbf{Quantum bit commitment using Wheeler's delayed choice experiment}}

\author{Chi-Yee Cheung}

\email{cheung@phys.sinica.edu.tw}

\affiliation{Institute of Physics, Academia Sinica\\
             Taipei, Taiwan 11529, Republic of China\\}

%\date{\today}

\begin{abstract}

We construct a quantum bit commitment scheme using a double-slit setup similar to Wheeler's delayed choice experiment. Bob sends photons toward the double-slit, and Alice commits by determining either the slit from which each photon emerges (for $b=0$), or its landing position on a screen (for $b=1$). Since the photon's wave front expands at the speed of light, Alice cannot delay the detection indefinitely, or it would very soon be out of her control no matter how much resources she has.

\end{abstract}



\maketitle

%%%%%%%%%%%%%%%%%%%%%% MAIN BODY %%%%%%%%%%%%%%%%%%%%%%%%%%%%%%

Quantum bit commitment is an important cryptographic primitive because
because it can be used as a building block of other more complex cryptographic protocols \cite{Brassard-96,Blum83,Bennett-91,Crepeau94,Yao95,
Mayers96,Kilian88,Crepeau-95}.
A bit commitment protocol involves two parties, Alice and Bob, who do not trust each other. Alice begins by committing to Bob
a secret bit $b\in\{0,1\}$ that is to be revealed to him at some
later time. In order to ensure Bob that she will keep her
commitment, Alice provides Bob with a piece of evidence
with which he can verify her honesty when she finally
unveils the committed bit. A bit commitment protocol is
secure if it satisfies the following two conditions. (1)
Concealing: Bob gets no information about the value of $b$
before Alice unveils it; (2) Binding: Alice cannot change
$b$ without Bob's knowledge. (In general both conditions
are satisfied only asymptotically as some security
parameter $N$ becomes large.) Furthermore, if the protocol
remains secure even if Alice and Bob had capabilities
limited only by the laws of nature, then it is said to be unconditionally secure.

As is well known, classical protocols cannot be unconditionally secure because their security always depends on some unfounded assumptions. In a quantum bit commitment (QBC) protocol, Alice and Bob execute a series of quantum and classical operations during the commitment procedure, which finally results in a quantum state $\rho_B^{(b)}$ in Bob's hand. If
 \beq
 \rho_B^{(0)}=\rho_B^{(1)}\label{perfect},
 \eeq
then the protocol is perfectly concealing, and Bob is not
able to extract any information about the value of $b$ from
$\rho_B^{(b)}$. When Alice reveals the value of $b$,
she must also provide additional information which,
together with $\rho_B^{(b)}$, will allow Bob to check if she is honest.

For some time a QBC protocol \cite{BCJL-93} proposed in
1993 was widely believed to be unconditionally secure, but
it was eventually shown otherwise \cite{Mayers95}.
Moreover, a ``no-go theorem" was put forth in 1997
\cite{LoChau97,Mayers97}, which supposedly proved that concealing
protocols can always be cheated by Alice, consequently
unconditionally secure QBC is ruled out as a matter
of principle. Due to the importance of QBC,
this result, if true, constitutes a major setback for
quantum cryptography.

Before proceeding further, it is instructive to outline the
impossibility proof for the perfectly concealing case \cite{LoChau97,Mayers97}. The proof is based on the assumption that, by quantum entanglement, Alice and Bob can
keep all undisclosed classical information undetermined and stored them at the quantum level. In other words, they can delay all prescribed
measurements without consequences until it is required to
disclose the outcomes. It then follows that at the end of the
commitment phase, there always exists a pure state
$\ket{\Psi^{(b)}_{AB}}$ in the joint Hilbert space
$H_A\otimes H_B$ of Alice and Bob. $\ket{\Psi^{(b)}_{AB}}$
is called a quantum purification of the evidence state
$\rho^{(b)}_B$, such that
 \beq
 {\rm Tr}_A~ \ket{\Psi^{(b)}_{AB}}
 \bra{\Psi^{(b)}_{AB}}
 =\rho_B^{(b)}, \label{reduced}
 \eeq
where the trace is over Alice's share of the state.
Note that whether Bob purifies or not is irrelevant to Alice, and she could simply assume he does.
In general purification requires access to
quantum computers, which is consistent with the assumption
that Alice and Bob have unlimited computational power.

Since the protocol is assumed to be perfectly concealing, therefore Eq. (\ref{perfect}) holds. Then, according to a theorem by Hughston \textit{et al.} \cite{Hughston-93}, the two purifications $\ket{\Psi^{(0)}_{AB}}$ and
$\ket{\Psi^{(1)}_{AB}}$ are related by a local unitary transformation $U_A$ on Alice's side, namely,
 \beq
 \ket{\Psi^{(1)}_{AB}} = U_A \ket{\Psi^{(0)}_{AB}}.
 \label{UA}
 \eeq
Since $U_A$ acts on $H_A$ only, she can implement it without Bob's help. Then it is obvious that Alice can change her commitment at will, and Bob will always conclude that she is honest because his density matrix $\rho_B^{(b)}$ is not affected by $U_A$. So it seems that no QBC protocols could be simultaneously concealing and binding.

It has been shown that, if we do not adhere to the standard
cryptographic scenario and allow Alice and Bob to control
two separated sites each, then an unconditionally secure
classical bit commitment protocol can be constructed using
the fact that signal transmission speed is finite \cite{Kent99}. In this paper, we are interested only in non-relativistic quantum bit commitment in the standard scenario where each party controls only one site.

Since 1997, there have been many attempts at breaking the
no-go theorem, however a mathematically
rigorous result is still lacking so far. Conceptually, the
claim that unconditionally secure QBC is ruled out in
principle is rather puzzling. First of all, while
the objective of QBC is well defined, the corresponding
procedure is not precisely specified. That is, there
exists no general mathematical characterization of all
imaginable QBC protocols \cite{Yuen08}. So it is
unlikely that the no-go theorem could have covered all
imaginable QBC protocols. Secondly,
when something in nature is forbidden, there usually exists
a deeper and more general reason or principle behind it.
Thus, for example, electric charge can neither be created
nor destroyed because electromagnetic interactions obey an
exact $U(1)$ gauge symmetry; arbitrary quantum states
cannot be perfectly cloned because information cannot be
transmitted faster than the speed of light; etc. However
unconditionally secure QBC is not known to violate any laws
in physics or information theory, so we believe that it is not strictly forbidden.

Notice that the impossibility proof does not consider the time evolution of the wavefunction involved. That is, the no-go theorem considers only stationary states with trivial time evolution properties which can be ignored. However a complete impossibility proof should also take into account the possible non-trivial time evolution of the quantum states involved. In this work, instead of stationary states such as ordinary qubits, we shall use photons whose spatial wavefunctions expand at the speed of light. We present below a practical protocol to illustrate our idea. The required experimental setup is a double-slit apparatus similar to Wheeler's delayed choice experiment, which is readily implementable.

\begin{itemize}
\item[]\hspace{-0.5cm}{Commit:}
\item[1.]{Bob sends an unpolarized photon toward a double-slit. With equal probability (1/3), the slits may be both open, or only one of which is open, or both closed. Alice is situated at a distance $L$ from the slits. She can detect the photon either on a screen or using telescopes focused on the slits. The photon enters the slits at $t=t_0$, and reaches Alice at $t=t_1$.}
\item[2.]{To commit to $b=0$, Alice uses the telescopes to determine the slit from which the photon emerges. For $b=1$, she detects the photon on the screen and records its landing position. Alice must announce whether a photon has been detected or not.}

\item[3.]{The above procedure is repeated $N$ times.}
\item[]\hspace{-0.5cm}{Unveil:}
\item[1.]{At time $t=T$, Alice unveils the value of $b$ and the corresponding detection data. Specifically, for $b=0$, she must specify the slit from which each of the photons went through; for $b=1$, she must reveal the position at which each photon landed on the screen.}
\item[2.]{Bob checks if Alice's data are consistent with her commitment.}
\end{itemize}

We now proceed to show that this protocol is secure.  It is obviously concealing, since the only information disclosed by Alice during the commitment procedure is if photons have been detected.

Next we show that it is binding. First of all, Alice must specify if a photon has reached her site, so she must try to detect it in one way or another. If so, it is clear that she can no longer change her commitment safely. If Alice determines the which-slit information, then she would not be able to distinguish the single-slit events from the double-slit ones, so that consistent reconstruction of the interference pattern on the screen would be impossible. On the other hand, if she detects the photons on the screen, the resulting pattern is a superposition of single-slit and double-slit events, and it is again impossible to separate the two because they occur randomly. Hence Alice cannot commit to one bit value and unveil another. If she did, she could only do so by pure guessing, but then her success probability ($P$) is only of order
\beq
P\sim 2^{-\frac{2}{3}N}
\eeq
which vanishes exponentially as $N\rightarrow\infty$.

Finally we consider quantum cheating strategy. As we saw, Alice has to determine whether a photon has gone through the double-slit and announce the result to Bob. This is in principle not difficult if she honestly detects it using either the screen or the telescopes. If she did, then clearly she cannot safely change her commitment later on, as explained earlier. Therefore, in order to be able to cheat later, Alice must detect the presence of the photon without disturbing its spatial wavefunction, which is however conceptually not possible because an unpolarized photon has no quantum properties other than its spatial wavefunction.

Even if Alice could somehow solve the detection problem, there is another insurmountable difficulty which we discuss below.
Let $\ket{\psi_{AB},t_1}$ be the wavefunction of the photon when it reaches Alice. It is not known how, but for the sake of discussion, let us suppose that Alice can somehow commit coherently to $b=0$, such that the resultant pure state can be schematically  written as
$
\ket{\Psi_{AB}(\vec x,t_1)}
=M_0\ket{\phi}\ket{\psi_{AB}(\vec x,t_1)}.
$
where $M_0$ is a unitary operator representing Alice's commitment action, and $\ket{\phi}$ stands for Alice's ancilla state. At $t=T$, when Alice is supposed to unveil, the total wavefunction becomes
\beq
\ket{\Psi_{AB}(\vec x,T)}
={\mathcal U}(T,t_1)M_0\ket{\phi}\ket{\psi_{AB}(\vec x,t_0)}.
\eeq
where ${\mathcal U}(T,t_1)=e^{-iH_p(T-t_1)}$
is the time evolution operator, and $H_p$ is the Hamiltonian of the photon. Note that without loss of generality we have ignored the time evolution of the ancilla state $\ket{\phi}$.

Now, in accordance with the no-go theorem, if at time $T$ Alice wants to safely change to $b=1$, she can apply a cheating unitary transformation $U_A$ on $\ket{\Psi_{AB}(\vec x,T)}$. However in the present case it is not possible to implement $U_A$ because the photon wave front grows at the speed of light and its size becomes astronomical within a short time. For example, even for a very modest commitment time of $\delta t=T-t_1=1$ second, the extent of the photon wavefunction is already many times the diameter of the earth. Theoretically Alice could still manipulate the wavefunction $\ket{\Psi_{AB}(\vec x,T)}$ if she had big enough instruments, but then she would not be able to do it in the privacy of her laboratory, and it would be no secret that she is cheating.
The above arguments show that Alice cannot cheat unless she could somehow stop the time evolution of the wavefunction $\ket{\Psi_{AB}^{(0)}(\vec x,t)}$, which is however not allowed by nature. Hence we conclude that this protocol is binding as well as concealing.

In summary, using a double-slit setup similar to Wheeler's delayed choice experiment, we have constructed a QBC protocol which is readily implementable. This protocol is secure because (1)It is not possible to detect the existence of an unpolarized photon without disturbing its spatial wavefunction; and (2)It is not possible to stop the time evolution of a photon wavefunction after emerging from the double-slit. Our result shows that non-trival time evolution of quantum states can be used to guarantee the security of QBC.

%%%%%%%%%%%%%%%%%%%%%% ACKNOWLEDGMENTS %%%%%%%%%%%%%%%%%%%%%%%%

%\acknowledgments{...}

%%%%%%%%%%%%%%%%%%%%%% REFERENCES %%%%%%%%%%%%%%%%%%%%%%%%%%%%%

%%%%%%%%%%%%%%%%%%%%%%%%%%%%%%%%%%%%%%%%%%%%%%%%%%%%%%%%%%%%%%%
\end{document}